\documentclass[twocolumn,twoside]{IEEEtran}

\IEEEoverridecommandlockouts

\usepackage{cite}
\usepackage{graphicx}
\usepackage{subfigure}
\usepackage{amsmath}
\usepackage{amssymb}
\usepackage{array}
\usepackage{multirow}
\usepackage{color,colortbl}
\usepackage{rotating}
\usepackage{booktabs}
\usepackage[
bookmarksopen,
bookmarksdepth=2,
breaklinks=true
]{hyperref}
\newcommand{\tabitem}{~~\llap{\textbullet}~~}

\begin{document}
\bibliographystyle{IEEEtran}
\title{Big Data Analytics for Manufacturing Internet of Things: Opportunities, Challenges and Enabling Technologies}


\author{
Hong-Ning Dai,\thanks{H.-N. Dai is with Faculty of Information Technology, Macau University of Science and Technology, Macau. E-mail: hndai@ieee.org.}
Hao Wang,\thanks{H. Wang is with Faculty of Engineering and Natural Sciences, Norwegian University of Science \& Technology, Gj{\o}vik, Norway. Email: hawa@ntnu.no}
Guangquan Xu,\thanks{G. Xu is with Tianjin Key Laboratory of Advanced Networking, 
), College of Intelligence and Computing, Tianjin University, Tianjin, China Email: losin@tju.edu.cn}
Jiafu Wan,\thanks{J. Wan is with School of Mechanical \& Automotive Engineering, 
South China University of Technology, Guangzhou, China Email: mejwan@scut.edu.cn}
Muhammad Imran\thanks{M. Imran is College of Computer and Information Sciences, King Saud University, Saudi Arabia,  Email: dr.m.imran@ieee.org}
}

\maketitle


\begin{abstract}
The recent advances in information and communication technology (ICT) have promoted the evolution of conventional computer-aided manufacturing industry to smart data-driven manufacturing. Data analytics in massive manufacturing data can extract huge business values while can also result in research challenges due to the heterogeneous data types, enormous volume and real-time velocity of manufacturing data. This paper provides an overview on big data analytics in manufacturing Internet of Things (MIoT). This paper first starts with a discussion on necessities and challenges of big data analytics in manufacturing data of MIoT. Then, the enabling technologies of big data analytics of manufacturing data are surveyed and discussed. Moreover, this paper also outlines the future directions in this promising area.

\end{abstract}

\begin{keywords}
Smart Manufacturing; Data Analytics; Data Mining; Internet of Things
\end{keywords}

\section{Introduction}
\label{sub:intro}

The manufacturing industry is experiencing a paradigm shift from automated manufacturing industry to ``smart manufacturing'' \cite{Kusiak:IJPR2018}. During this evolution, Internet of Things (IoT) plays an important role of connecting the physical environment of manufacturing to the cyberspace of computing platforms and decision-making algorithms, consequently forming a Cyber-Physical System (CPS) \cite{hndai:BDAWireless2019}. We name such industrial IoT dedicated to manufacturing industry as manufacturing IoT (MIoT) in this paper.

MIoT consists of a wide diversity of manufacturing equipments, sensors, actuators, controllers, RFID tags and smart meters, which are connected with computing platforms through wired or wireless communication links. There is a surge of big volume of data traffic generated from MIoT. The MIoT data is featured with large volume, heterogeneous types (i.e., structured, semi-structured, unstructured) and is generated in a real-time fashion. The analytics of MIoT data can bring many benefits, such as improving factory operation and production, reducing machine downtime, improving product quality, enhancing supply chain efficiency and improving customer experience \cite{RayZhong:IJPR2017,lade2017manufacturing,TAO:2018}. However, there are also many challenges in data analytics in MIoT in the different phases of the whole life cycle of data analytics. 

There are several surveys on data analytics in manufacturing industry. The work of \cite{TAO:2018} proposes a data-driven smart manufacturing framework and provides several application scenarios based on this conceptual framework. The necessities of big data analytics in smart manufacturing are summaried in \cite{kusiak2017smart}. The work of \cite{lade2017manufacturing} provides an overview on data analytics in manufacturing with a case study. Tao and Qi presents an overview of service-oriented manufacturing in \cite{FTao:TSMC19}. However, most of the aforementioned studies lack of the introduction of enabling technologies corresponding to the challenges, which are of interest to both academic researchers and industrial practitioners.

Therefore, the aim of this paper is to provide an overview on data analytics in MIoT from opportunities, challenges and enabling technologies. The main contributions of this paper can be summarized as follows.
\begin{itemize}
\item We provide a summary on key characteristics of MIoT and a life cycle of big data analytics for MIoT data. We also discuss necessities and challenges of big data analytics in MIoT.
\item We present an overview on enabling technologies of big data analytics for MIoT from the aspects of data acquisition, data preprocessing and data analytics.
\item We given an outline of future research directions in aspects of security, privacy, fog computing and new data analytics methods. 
\end{itemize}

The rest of this paper is organized as follows. Section \ref{sec:nec-challenges} gives the discussion on necessities and challenges of big data analytics in MIoT. Section \ref{sec:enabling} introduces enabling technologies of big data analytics in MIoT. Section \ref{sec:open} discusses the future research directions. Finally, this paper is concluded in Section \ref{sec:conc}.



%
%
%

\section{Necessities and challenges of big data analytics for Manufacturing Internet of Things}
\label{sec:nec-challenges}

In this section, we first introduce the key characteristics of Manufacturing Internet of Things in Section \ref{subsec:key-IIoT}. We then introduce the life cycle of big data analytics for MIoT in Section \ref{subsec:lifecycle}. We next discuss the necessities of big data analytics for MIoT in Section \ref{subsec:necessities} and the challenges in Section \ref{subsec:challenges}.

\subsection{Key characteristics of Manufacturing Internet of Things}
\label{subsec:key-IIoT}

In this paper, we roughly categorize IoT into consumer Internet of Things (CIoT) and Manufacturing Internet of Things (MIoT). Table \ref{tab:IIoTvsCIoT} compares MIoT with CIoT. In contrast to MIoT, CIoT mainly serve for consumers. Hence, CIoT mainly consists of  consumer devices (e.g., smart phones, wearable electronics) and smart appliances (e.g., refrigerators, TVs, washing machines). CIoT mainly aims to improve user experience while MIoT mainly focuses on improving factory operations and production, reducing the machine downtime and improving product quality. Moreover, MIoT usually works in harsh industrial environment (like vibrated, noisy and extremely high/low temperature) while CIoT works in moderate environment. In addition, MIoT applications usually require high data-rate network connection with low delay while CIoT applications have relaxed requirement on network connection. Furthermore, MIoT systems are usually mission-critical and sensitive to system failure or machinery downtime while CIoT systems are non-mission-critical. 

In this paper, we mainly focus on MIoT. The MIoT ensures the connection of various \emph{things} (smart objects) mounted with various electronic or mechanic sensors, actuators, instruments and software systems which can sense and collect information from the physical environment and then make actions on the physical environment. During this procedure, the data analytics plays an important role in extracting informative values, forecasting the coming events and predicting the increment/decrements of products.

\begin{table*}[t]
\caption{Comparison between MIoT and CIoT}
\centering
\renewcommand{\arraystretch}{2.5}
\begin{tabular}{m{2.3cm}|m{5cm}|m{5cm}}
\hline
& \textbf{Manufacturing IoT} & \textbf{Consumer IoT} \\
\hline
\hline
Goal & Manufacturing-industry Centric & Consumer Centric\\
\hline
Devices & Machines, Sensors, Controllers, Actuators, Smart meters & Consumer devices and Smart appliances\\
\hline
Working Environment & Harsh (vibration, noisy, extremely high/low temperature) & Moderate\\
\hline
Data rate & High (usually) & Low or average \\
\hline 
Delay & Delay sensitive & Delay tolerant \\
\hline
Mission & Mission-critical & Non-mission-critical\\
\hline
\end{tabular}
\label{tab:IIoTvsCIoT}
\end{table*}

\subsection{Life cycle of big data analytics for MIoT}
\label{subsec:lifecycle}

We first introduce the life cycle of big data analytics for MIoT. Figure \ref{fig:BDA-lifecycle} shows that the life cycle of big data analytics for MIoT consists of three consecutive stages: 1) Data Acquisition, 2) Data Preprocessing and Storage, 3) Data Analytics. There are other taxonomies \cite{Hu:IEEEAccess14,Casado:CCPE2015,TAO:2018}.  We categorize the life cycle of big data analytics into the above three stages since this taxonomy can accurately capture the key features of big data analytics in MIoT.

\begin{enumerate}
\item[1.] \emph{Data acquisition} consists of data collection and data transmission. Firstly, data collection involves acquiring raw data from various data sources in the whole manufacturing process via dedicated data collection technologies. For example, RFID tags are scanned by RFID readers in product warehouse. Then, the collected data will be transmitted to the data storage system through either wired or wireless communication systems. Details about enabling technologies of data acquisition are given in Section \ref{subsec:communication}.

\item[2.] \emph{Data preprocessing and storage.} After data collection, the raw data needs to be preprocessed before keeping them in data storage systems because of the big volume, redundancy, uncertainty features of the raw data \cite{lade2017manufacturing}. The typical data preprocessing techniques include data cleaning, data integration and data compression. Data storage refers to the process of storing and managing massive data sets. We divide the data storage system into two components: storage infrastructure and data management software. The infrastructure not only includes the storage devices but also the network devices connecting the storage devices together. In addition to the networked storage devices, data management software is also necessary to the data storage system. Details about enabling technologies of data preprocessing and data storage are given in Section \ref{subsec:storage}.

\item[3.] \emph{Data analytics.} In data analysis phase, various data analytical schemes are used to extract valuable information from the massive manufacturing data sets. We roughly categorize the data analytical schemes into four types: (i) statistic modelling, (ii) data visualization, (iii) data mining and (iv) machine learning. Details about enabling technologies of data analysis are presented in Section \ref{subsec:analytics}.
\end{enumerate}

\subsection{Necessities of big data analytics for MIoT}
\label{subsec:necessities}

\begin{figure*}[t]
\centering
\includegraphics[width=14cm]{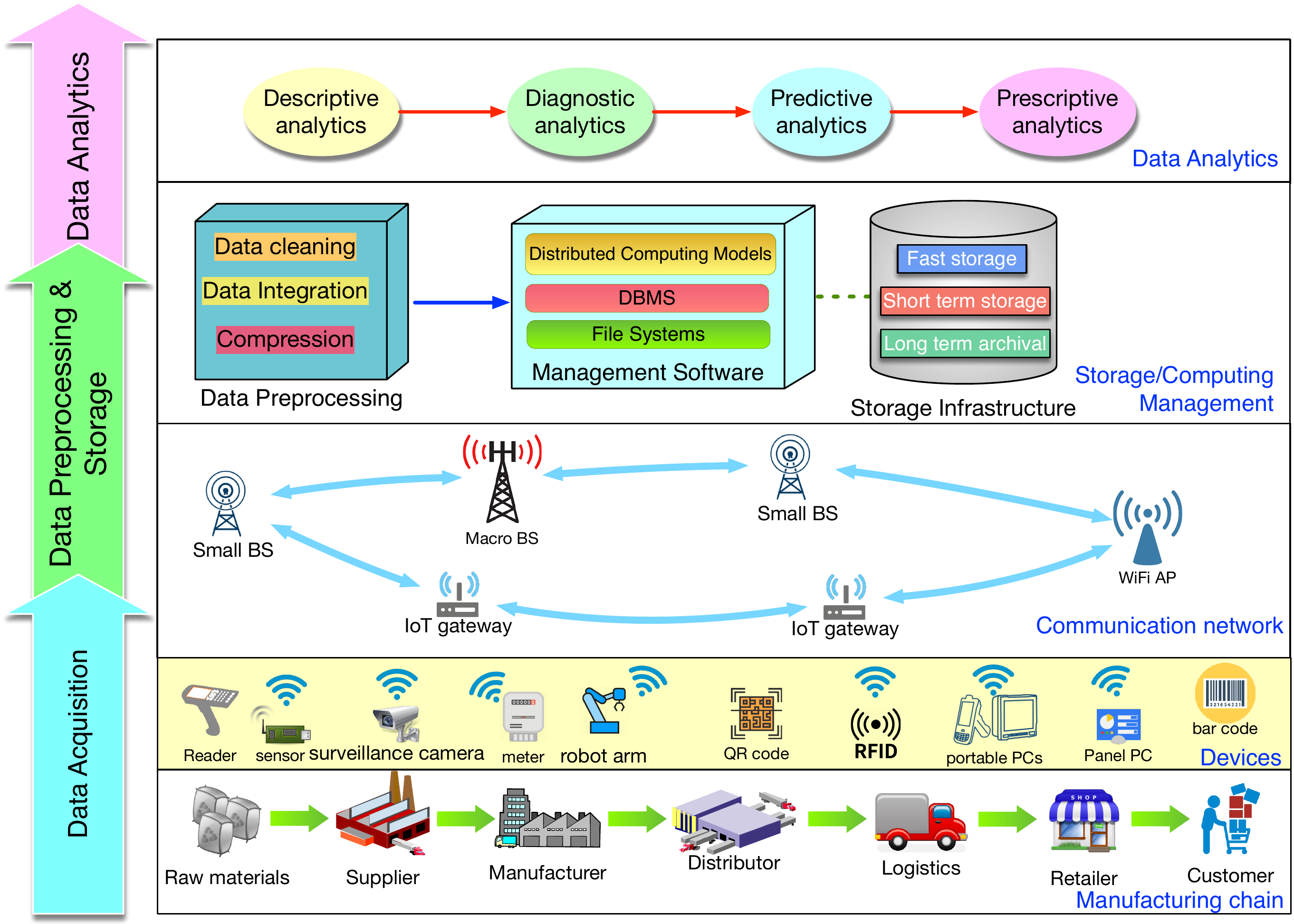}
\caption{Life cycle of Big Data Analytics for MIoT}
\label{fig:BDA-lifecycle}
\end{figure*}

There is an enormous amount of data generated from the whole manufacturing chain consisting of raw material supply, manufacturing, product distribution, logistics and customer support, as shown in Figure \ref{fig:BDA-lifecycle}. Such ``big data'' needs to be extensively analysed so that some valuable and informative information can be extracted.

We summarize the reasons of big data analytics for MIoT as follows:
\begin{itemize}

\item \emph{Improving factory operations and production.} The predictive analytics of manufacturing data and customer demand data can help to improve machinery utilization consequently enhancing factory operations. For example, the demands for certain products are often related to weather or seasonal conditions (e.g., down coats related to the cold weather). Forecasting a cold wave can be used to make pro-active allocation of machinery resources and pre-purchasing raw materials to fulfill the upsurge demands. 

\item \emph{Reducing machine downtime.} The prevalent sensors deployed throughout the whole product line can collect various data reflecting machinery status. For example, the analysis of machinery health data can help to identify the root cause of failure consequently reducing machine downtime \cite{lade2017manufacturing}. Moreover, the sensory data from automatic assembly line can also be used to determine excessive load of machines so as to balance the loads among multiple machines \cite{wang2018deep}.

\item \emph{Improving product quality.} On one hand, the analysis of market demand and customer requirement can be used to improve the product design in reflecting product improvements. During the product manufacturing procedure,  the analysis of manufacturing data can help to reduce the ratio of defective goods by identifying the root cause. As a result, the product quality can be improved. 

\item \emph{Enhancing supply chain efficiency.} The proliferation of various sensors, RFID and tags during supplier, manufacturing and transportation generates massive supply chain data, which can be used to analyse supply risk, predict delivery time, plan optimal logistic route, etc. Moreover, the analysis of inventory data can reduce the holding costs and fulfill the dynamic demands by establishing safety stock levels. In addition, big data analytics on IoT-enabled intelligent manufacturing shops \cite{RayZhong:IJPR2017} can also help to make accurate logistic plan and schedules. As a result, the system efficiency can be greatly improved.

\item \emph{Improving customer experience.} Companies can obtain customer data from various sources, such as sales channels, partner distributors, retailers, social media platforms. Then, big data analytics on customer data offers descriptive, predictive and prescriptive solutions to enable companies to improve product design, quality, delivery, warrant and after-sales support. As a result, the customer experience can be improved. For example, the IoT data in the whole food supply-chain is also beneficial to prevent mischievous actions and guarantee food safety \cite{Leng2018}.
\end{itemize}

\subsection{Challenges of big data analytics for MIoT}
\label{subsec:challenges}

MIoT data has the following characteristics: (1) massive volume, (2) heterogeneous data types, (3) being generated in real-time fashion and (4) bringing huge both business value and social value. The unique features cause the research challenges in big data analytics for MIoT. We summarize the challenges in the following aspects.

\emph{1. Challenges in data acquisition}

Data acquisition addresses the issues including data collection and data transmission, during which there are the following challenges.

\begin{itemize}

\item \emph{Difficulty in data representation}. MIoT data has different types, heterogeneous structures and various dimensions. For example, manufacturing data can be categorized into structured data, semi-structured and un-structured data \cite{TAO:2018}. How to represent these structured, semi-structured and un-structured data becomes one of major challenges in big data analytics for MIoT.

\item \emph{Efficient data transmission}. How to transmit the tremendous volumes of data to data storage infrastructure in an efficient way becomes a challenge due to the following reasons: (i) \emph{high bandwidth consumption} since the transmission of big data becomes a major bottleneck of wireless communication systems \cite{Hu:IEEEAccess14}; (ii) \emph{energy efficiency} is one of major constraints in many wireless industrial systems, such as industrial wireless sensor networks \cite{Azoidou:TII2017}.

\end{itemize}

\emph{2. Challenges in data preprocessing and storage}

Data generated from MIoT leads to the following research challenges in data preprocessing.

\begin{itemize}

\item \emph{Data integration}. Data generated in MIoT has the various types and heterogeneous features. It is necessary to integrate the various types of data so that efficient data analytics schemes can be implemented. However, it is quite challenging to integrate different types of MIoT data. 

\item \emph{Redundancy reduction}. The raw data generated from MIoT is characterized by the temporal and spatial redundancy, which often results in the data inconsistency consequently affecting the subsequent data analysis. How to mitigate the data redundancy in MIoT data becomes a challenge.  

\item \emph{Data cleaning and data compression}. In addition to data redundancy, MIoT data is often erroneous and noisy due to the defected machinery or errors of sensors. However, the large volume of the data makes the process of data cleaning more challenging. Therefore, it is necessary to design effective schemes to compress MIoT data and clean the errors of MIoT data.  
\end{itemize}

Data storage plays an important role in data analysis and value extraction. However, designing an efficient and scalable data storage system is challenging in MIoT. We summarize the challenges in data storage as follows.

\begin{itemize}

\item \emph{Reliability and persistency of data storage}. Data storage systems must ensure the reliability and the persistency of MIoT data. However, it is challenging to fulfill the above requirements of big data analytics while balancing the cost due to the tremendous amount of data \cite{Guerra:FAST2011}. 

\item \emph{Scalability}. Besides the storage reliability, another challenging issue lies in the scalability of storage systems for big data analytics. The various data types, the heterogeneous structures and the large volume of massive data sets of MIoT lead to the in-feasibility of conventional databases in big data analytics. As a result, new storage paradigms need to be proposed to support large scale data storage systems for big data analytics.

\item \emph{Efficiency}. Another concern with data storage systems is the efficiency. In order to support the vast number of concurrent accesses or queries initiated during the data analytics phase, data storage needs to fulfill the efficiency, the reliability and the scalability requirements together, which is extremely challenging.

\end{itemize}

\emph{3. Challenges in data analytics}

It is quite challenging in big data analytics for MIoT due to the tremendous volume, the heterogeneous structures and the high dimension. The major challenges in this phase are summarized as follows.

\begin{itemize}
\item \emph{Data temporal and spatial correlation}. Different from conventional data warehouses, MIoT data is usually spatially and temporally correlated. How to manage the data and extract valuable information from the temporally/ spatially-correlated MIoT data becomes a new challenge. 

\item \emph{Efficient data mining schemes}. The tremendous volume of MIoT data leads to the challenge in designing efficient data mining schemes due to the following reasons: (i) it is not feasible to apply conventional multi-pass data mining schemes due to the huge volume of data, (ii) it is critical to mitigate the data errors and uncertainty due to the erroneous features of MIoT data.

\item \emph{Privacy and security}. It is quite challenging to pertain the privacy and ensure the security of data during the analytics process. Though there are a number of conventional privacy-preserving data analytical schemes, they may not be applicable to the MIoT data with the huge volume, heterogeneous structures, and spatio-temporal correlations. Therefore, new privacy-preserving data mining schemes need to be proposed to address the above issues.

\end{itemize}

\section{Enabling Technologies}
\label{sec:enabling}

In this section, we discuss the enabling technologies of big data analytics in MIoT. According to the three phases in the life cycle of big data analytics in MIoT, we roughly categorize these technologies into data acquisition, data preprocessing and storage, data analytics. In particular, we first discuss the data acquisition related technologies in Section \ref{subsec:communication}. We then describe the data preprocessing and storage in Section \ref{subsec:storage}. In Section \ref{subsec:analytics}, we discuss the data analytics in MIoT.

\subsection{Data acquisition}
\label{subsec:communication}

\begin{figure}[t]
\centering
\includegraphics[width=8cm]{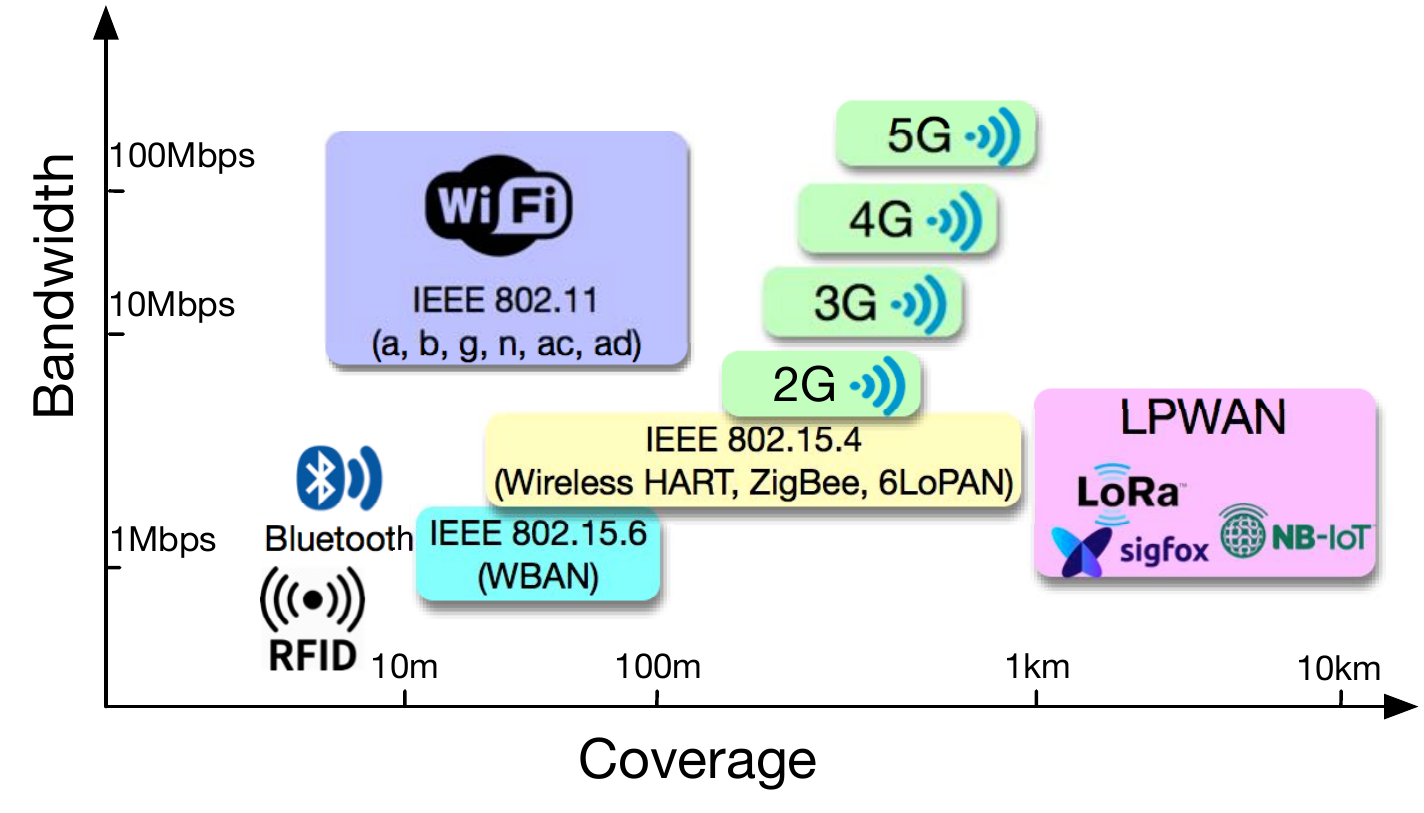}
\caption{Wireless Communication Technologies for MIoT (figure is not scalable)}
\label{fig:Wireless}
\end{figure}

As shown in Figure \ref{fig:BDA-lifecycle}, the whole manufacturing chain involves with multiple parties such as suppliers, manufacturers, distributors, logistics, retailers and customers. As a result, different types of data sources generate from each of these sectors. Take a manufacturing factory an example. Sensors deployed at the production line can collect device data, product data, ambient data (like temperature, humidity, air pressure), electricity consumption, etc. In the product warehouse, RFID or other tags can help to identify and track products. RFID tags attached at products can be read in a short distance by a RFID reader in a wireless manner.

The collected data can then be transmitted to the next stage via either wired or wireless manner. Industrial Ethernet is one of the most typical wired connections in manufacturing. When Ethernet is applied to an industrial setting, more rugged connectors and more durable cables are often required to satisfy harsh environment requirements (like vibration, noise and temperature). Compared with wired communications, wireless communications do not require communication wiring and related infrastructure consequently saving the cost and improving scalability. The major obstacle of the wide deployment of wireless communications in industrial systems is the lower throughput and the higher delay than wired communications. However, the recent advances in wireless communications make wireless connections feasible in industrial components.

Various sensors, RFIDs and other tags can connect with IoT gateways, WiFi Access Points (APs), small base station (BS) and macro BS to form an industrial wireless sensor networks (IWSN) \cite{QChi:TII14}. It is worth mentioning that different wireless technologies have different coverage and bandwidth capabilities. Figure \ref{fig:Wireless} gives the comparison of various wireless technologies regarding to coverage and bandwidth. In particular, it is shown in Figure \ref{fig:Wireless} that conventional wireless technologies like Near Field Communications (NFC), RFID, Bluetooth Low Energy (LE), wireless body sensor networks (WBAN), Internet Protocol (IPv6), Low-power Wireless Personal Area Networks (6LoWPAN) and Wireless Highway Addressable Remote Transducer (WirelessHART) \cite{Petersen:IEEE2011} are suffering from short communication range (i.e., most of them can typically cover less than hundreds of meters). As a result, they cannot support the wide-coverage industrial applications, like smart metering, smart cities and smart grids \cite{JXu:IOTJ2017}. It is true that other wireless technologies such as WiFi (IEEE 802.11) and mobile communication technologies (such as 2G, 3G, 4G networks) can provide longer coverage range while they often require high energy consumption at handsets, whereas most of sensor nodes have the limited energy (i.e., supplied by batteries). Therefore, WiFi and other mobile communication technologies may not be feasible in IWSN due to the high energy consumption.

Recently, Low Power Wide Area Networks (LPWAN) essentially provide a solution to the wide coverage demand while saving energy. Typically LPWAN technologies include Sigfox, LoRa, Narrowband IoT (NB-IoT) \cite{MEKKI2018}. LPWAN has lower power consumption than WiFi and mobile communication technologies. Take NB-IoT as an example. It is shown in \cite{JXu:IOTJ2017} that an NB-IoT node can have a ten-year battery life. Moreover, LPWAN has a longer communication range than RFID, bluetooth and 6LoWPAN. In particular, LPWAN technologies have the communication range from 1km to 10 km. Furthermore, they can also support a large number of concurrent connections (e.g., NB-IoT can support 52,547 connections as shown in \cite{JXu:IOTJ2017}). However, one of limitations of LPWAN technologies is the low data rate (e.g., NB-IoT can only support a data rate upto 250 kps). Therefore, LPWAN technologies should complement with conventional RFID, 6LoWPAN and other wireless technologies so that they can support the various data acquisition requirements. 

%
%

\subsection{Data preprocessing and storage}
\label{subsec:storage}

\subsubsection{Data preprocessing}

Data acquired from MIoT has the following characteristics:

\begin{itemize}

\item \emph{Heterogeneous data types}. The whole manufacturing chain generates various data types including sensory data, RFID readings, product records, text, logs, audio, video, etc. The data is in the forms of structured, semi-structured and non-structured. 

\item \emph{Erroneous and noisy data}. The data obtained from industrial environment is often erroneous and noisy mainly due to the following reasons: (a) interference during the process of data collection especially in industrial environment, (b) the failure and malfunction of sensors or machinery, (c) intermittent loss or outage of wireless or wired communications \cite{SIDDIQA:JNCA2016}. For example, wireless communications are often susceptible to harsh industrial environmental factors like blockage, shadowing and fading effects. Moreover, data transmission may fail in industrial WSNs due to the depletion of batteries of sensors or machinery.

\item \emph{Data redundancy}. Data generated in MIoT often contain excessively redundant information. For instance, it is shown in \cite{Ertek:TSMCS2017} that there are excessive duplicated RFID readings when multiple RFID tags were scanned by several RFID readers at different time slots. The data redundancy often results in data inconsistency. 

\end{itemize}

\begin{figure}[t]
\centering
\includegraphics[width=8cm]{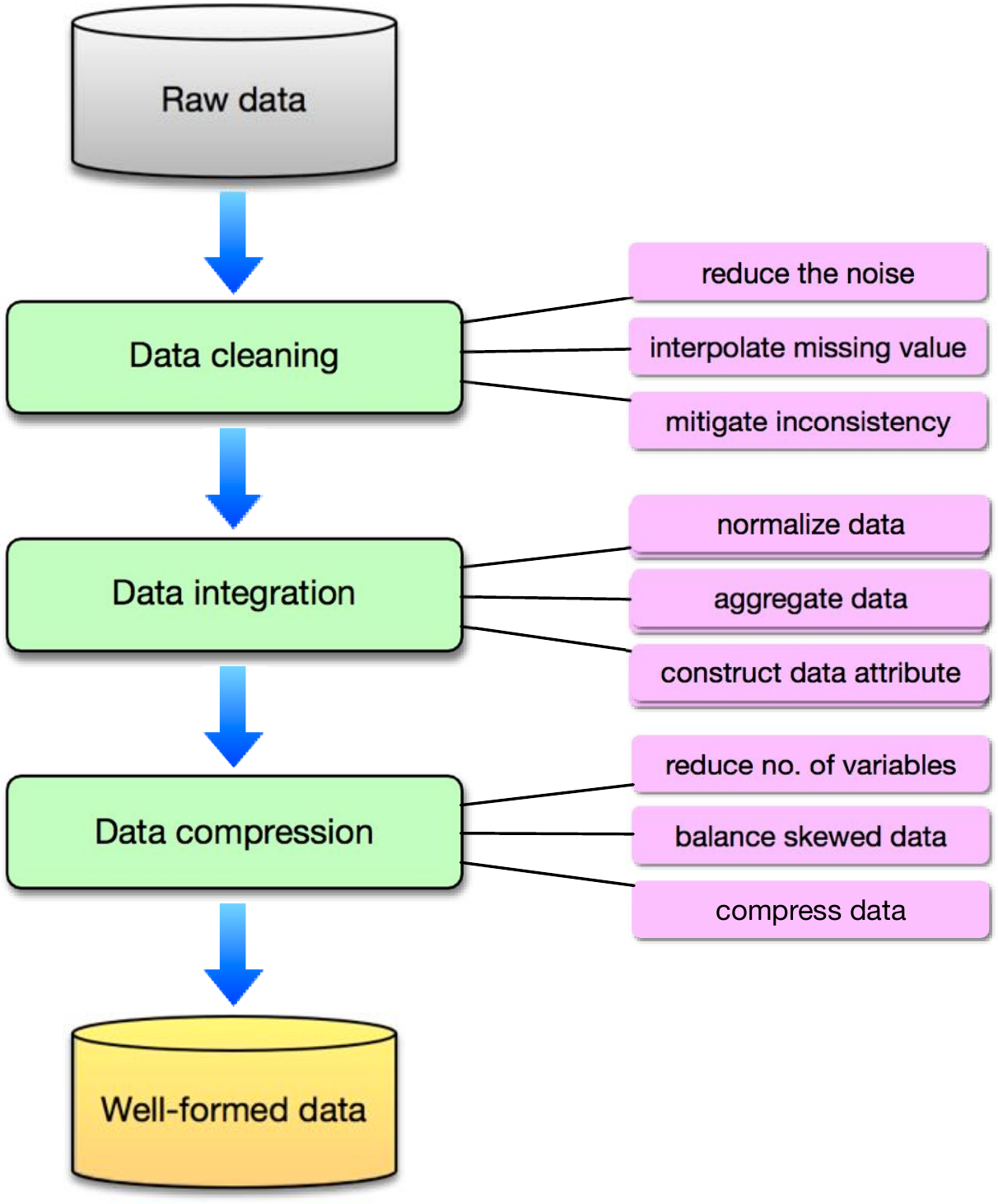}
\caption{Data preprocessing techniques}
\label{fig:preprocessing}
\end{figure}

Data preprocessing approaches on MIoT data include \emph{data cleaning}, \emph{data integration} and \emph{data compression} as shown in Figure \ref{fig:preprocessing}. In industrial environment, sensory data is usually uncertain and erroneous due to the depletion of battery power of sensors, imprecise measurement of sensors and communication failures. There are several approaches proposed to address these issues. For example, \cite{ZHONG:2015} proposed RFID-Cuboids approach to remove redundant readings and eliminate the missing values. Moreover, an Indoor RFID Multi-variate Hidden Markov Model (IR-MHMM) was proposed to determine uncertain data and remove duplicated RFID readings as shown in \cite{Baba:2017}. Furthermore, a machine-learning based method was proposed to filter out the invalid RFID readings \cite{MA:ESA2018}. In addition, the study of \cite{Bhandari:sensors17} proposed an auto-correlation based scheme to remove duplicated time-series temperature data. In \cite{Tasnim:CCNC17}, a novel data cleaning mechanism was proposed to clean erroneous data in environmental sensing applications. Besides duplicated readings, there also exist missing values in MIoT data. In \cite{ZZheng:TII18}, an interpolation method was proposed to recover the missing values of smart grids data. Moreover, energy-saving is a critical issue in data-cleaning algorithms used in MIoT. In \cite{CDeng:IJPR2018}, an energy-efficient data-cleaning scheme was proposed. 

\subsubsection{Data storage}

Data storage plays an important role in big data analytics for MIoT. We summarize the solutions of data storage in two aspects: 1) storage infrastructure and 2) data management software. 

\emph{Storage infrastructure} consists of a number of interconnected storage devices. Storage devices typically include: magnetic Harddisk Drive, Solid-State Drives, magnetic taps, USB flash drives, Secure Digital (SD) cards, micro SD cards, Read-Only-Memory (ROM), CD-ROMs, DVD-ROMs, etc. These storage devices can be connected together (via wired or wireless connections) to form the storage infrastructure for MIoT in industrial environment. 

Besides storage infrastructure, \emph{data management software} plays an important role in constructing the scalable, effective, reliable storage system to support big data analytics in MIoT. As shown in Figure \ref{fig:BDA-lifecycle}, the data management software consists of three layered components: 

\begin{itemize}
\item \textit{Distributed file systems.} Google File System (GFS) was proposed and developed by Google \cite{Ghemawat:GFS2003} to support the large data intensive distributed applications such as search engine. Moreover, Hadoop Distributed File System (HDFS) was proposed by Apache  \cite{Shvachko:2010} as an alternative to GFS. In addition, there are other distributed file systems, such as C\# Open Source Managed Operating System (Cosmos) proposed by Microsoft \cite{Chaiken:2008}, XtreemFS \cite{Hupfeld:2008} and Haystack proposed by Facebook \cite{Beaver:2010}. Most of them can partially or fully support the storage of large scale data sets. Therefore, most of them can offer the support for large scale data storage of MIoT data.

\item \textit{Database management systems (DBMS).} DBMS offers a solution to organize the data in an efficient and effective manner. DBMS software tools can be roughly categorized into two types: traditional relational DBMS (aka SQL databases) and non-relational DBMS (aka Non-SQL databases). SQL databases have been a primary data management approach, especially useful to Material Requirements Planning (MRP), Supply Chain Management (SCM), Enterprise Resource Planning (ERP) in the whole manufacturing chain. Typical SQL databases including commercial databases, such as Oracle, Microsoft SQL server and IBM DB2, and open-source alternatives, such as MySQL, PostgreSQL and SQLite. SQL databases usually store data in tables of records (or rows). This storage method neverthless leads to the poor scalability of databases. For example, when data grows, it is necessary to distribute the load among multiple servers. One of benefits of SQL databases is that most of SQL databases can guarantee ACID (Atomicity, Consistency, Isolation, Durability) properties of database transactions, which is crucial to many commercial applications (e.g., ERP and inventory management). Different from SQL databases, NoSQL databases support various types of data, such as records, text, and binary objects. Compared with traditional relational databases, most of NoSQL databases are usually highly scalable and can support the tremendous amount of data. Therefore, NoSQL databases are promising in managing sensory data, device data, RFID trajectory data in MIoT \cite{lade2017manufacturing}.

\item \textit{Distributed computing models}. There are a number of distributed computing models proposed for big data analytics. For example, Google MapReduce \cite{Dean:2008} is one of the typical programming models used for processing large data sets. Hadoop MapReduce \cite{Hadoop:MapReduce} is the open source implementation of Google MapReduce. MapReduce is suffering from the lack of iterations or recursions, which are however required by many data analytics applications, such as data mining, graph analysis and social network analysis. There are some extensions to MapReduce to address this concern, including HaLoop \cite{Bu:VLDB2010}, Berkeley Orders of Magnitude (BOOM) Analysis \cite{Alvaro:2010}, Twister \cite{Ekanayake:2010}, iHadoop \cite{Elnikety:CloudCom11} and iMapReduce \cite{Zhang:2012}. In addition to MapReduce, there are other alternatives such as Dryad \cite{Isard:2007}, Nephele/PACTs system \cite{Battre:socc2010}, Spark \cite{Zaharia:2010}, Pregel \cite{Malewicz:2010}, Hive \cite{Thusoo:icde2010}, GraphLab \cite{Low:2012}.

\item \textit{Virtual machines and containers}. Virtual machines (VMs) have been widely used to support cloud computing. Through virtualization, multiple VMs can be emulated on a single computer system. VMs can help to achieve the isolation of multiple virtual operating systems, on top of which multiple applications can be supported. Different from VMs, containers run on top of a single operating system and a single hardware while containers separate the applications as well as the underneath binary and library files. Therefore, containers can achieve the lightweight virtualization, consequently resulting the super fast booting speed, small size, less resource consumption (compared with VMs). The lightweight features of containers lead to the feasibility to edge computing scenarios (to be illustrated in Section \ref{subsec:case}).  
\end{itemize}

\subsection{Data analytics}
\label{subsec:analytics}

\subsubsection{Typical data analytics approaches}

Typical data analytics approaches include: 1) \emph{Statistical modeling} schemes, 2) \emph{Data mining} schemes, 3) \emph{Machine learning} schemes and 4) \emph{Data visualization}. 

Statistical modeling methods are mainly based on statistical theory. There are three types of statistical methods: (i) descriptive statistics that is used to quantify relationships in data \cite{Trochim:2016}; (ii) inferential statistics that is used to to deduce generalizations from the sample data sets \cite{Bandyopadhyay:2011}; (iii) stochastic modeling methods can capture the dynamic features of data traffic, predict user mobility and track objects \cite{Newson:2009,YLiao:EIS18}.

Data mining is the process of extracting useful information from massive data sets. There are a wide variety of data mining algorithms that can be used in MIoT such as Apriori algorithm, Frequent Pattern Growth (FP-Growth) algorithm, Density-based spatial clustering of applications with noise (DBSCAN), Generalized Sequential Pattern (GSP), Sequential Pattern Discovery Using Equivalent Class (SPADE) and Prefix-Projected Sequential Pattern Mining (PrefixSpan)
\cite{Han:2012}.

Machine learning explores to construct self-adaptive algorithms that can learn from existing data and perform predictive analysis. As one of typical applications of machine learning, data mining has emphasis on extracting valuable information from data. Typical Machine learning algorithms include support vector machines (SVMs) \cite{Vapnik:1995}, naive Bayes \cite{Wu:2007}, Decision tree learning \cite{Russell:2009}, $k$-Nearest Neighbors ($k$-NN) \cite{altman:1992}, hidden Markov model, Bayesian networks \cite{Qiu:2016}, neural networks \cite{GPZhang:2000}, Ensemble methods \cite{Zhou:2012}, $k$-means \cite{Kanungo:2002}, singular value decomposition (SVD), Principal Component Analysis (PCA) \cite{Jolliffe:2002} and reinforcement learning algorithms such as Q-learning \cite{Russell:2009}.

\begin{figure*}[t]
\centering
\includegraphics[width=12cm]{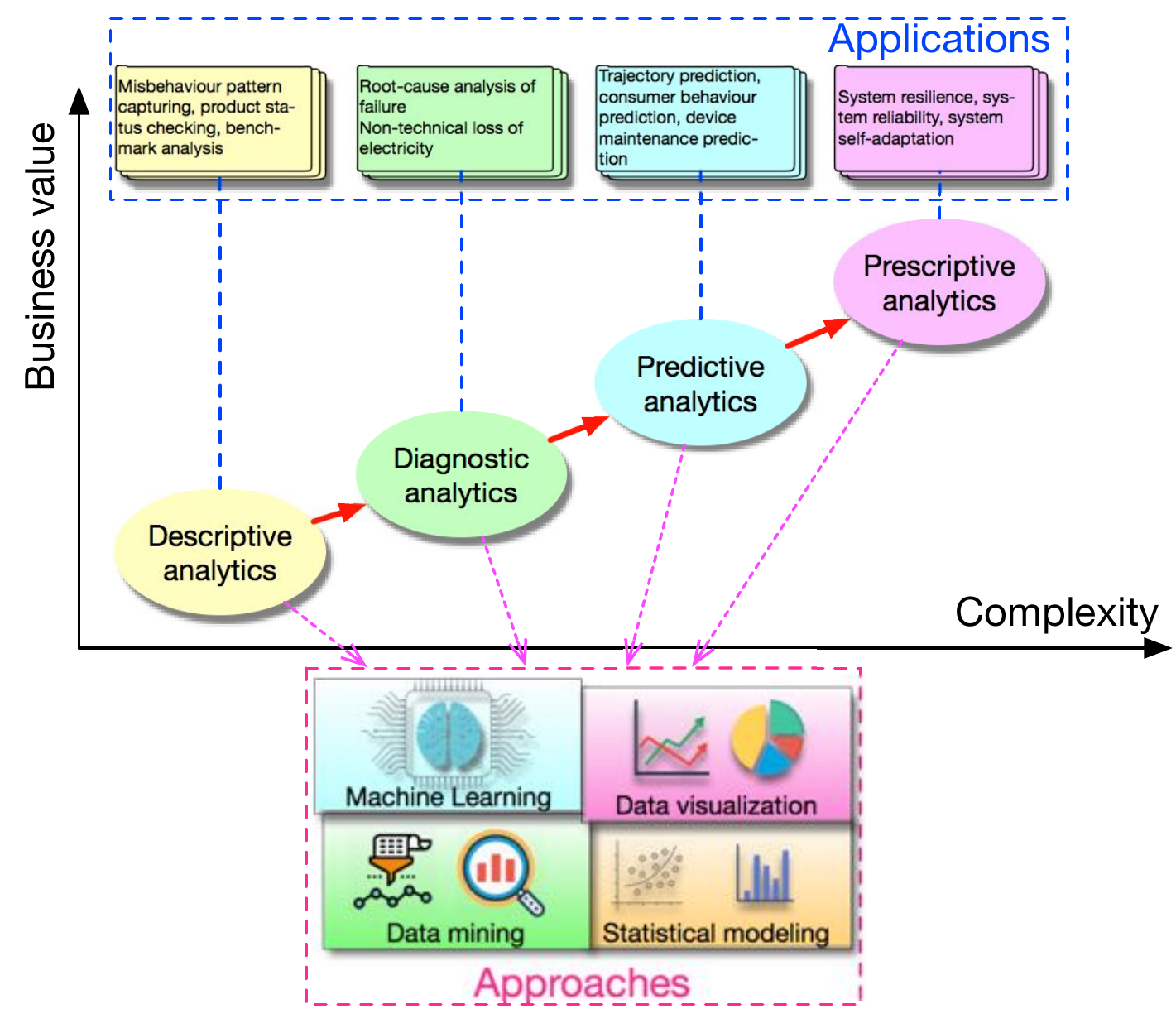}
\caption{Data analytics}
\label{fig:da}
\end{figure*}

\subsubsection{Taxonomy of data analytics approaches in MIoT}

We next present an overview of data analytics in MIoT in the aspect of MIoT applications. In particular, data analytics methods in MIoT can be roughly categorized into: 1) Descriptive analytics, 2) Diagnostic analytics, 3) Predictive analytics, 4) Prescriptive analytics. This classification can better represent the data analytics in MIoT applications in different levels of complexity and extracted values. Figure \ref{fig:da} depicts different levels of data analytics methods in MIoT applications. Both descriptive and diagnostic analytics methods are reactive while predictive and prescriptive analytics approaches are proactive. Moreover, prescriptive and predictive analytics approaches are more complicated than descriptive and diagnostic analytics methods though they can bring more values than descriptive and diagnostic analytics. We then present an overview of existing studies in the four levels of data analytics. 

\emph{(1) Descriptive analytics} 

Descriptive analytics is an exploratory analysis of historical data to tell what happened. During this stage, most of data mining and statistic methods can be used to reveal the data characteristics, recognize patterns and identify relationships of data objects. Descriptive analytics can be used in the whole life cycle of manufacturing data. In particular, a real-time monitoring system was proposed in \cite{zhang2015:real} to track the different manufacturing resources. Zhong et al. \cite{Zhong:IJAMT2016}  proposed RFID-Cuboid framework to integrate production logistic data with RFID data and offered a system prototype to visualize logistic trajectory data. Moreover, the study of \cite{zuo2018internet} presented a cloud-based approach to evaluate the energy consumption during product manufacturing process. In addition, air-qualtiy monitoring system based on wireless sensor networks at a logistics shipping base was proposed in \cite{MOLKADANIELSEN:2018}.

\emph{(2) Diagnostic analytics} 

Diagnostic analytics is a deeper look at data to attempt to understand the causes of events and behaviours. The diagnostic analysis of machines and other equipments can help to identify the possible faults and predict the failures to reduce the machine down-times. For example, a method of integrating SVM and artificial neural network (ANN)  was presented to detect and diagnose machinery faults of centrifugal pumps \cite{AZADEH:2013}. The study of \cite{HWang:OCEANS2016} proposed fault detection methods for propeller ventilation of vessels based on Kalman filter. Wuest et al. put forth a surpervised maching learning method to monitor product quality in \cite{Wuest:2014}. Compared with supervised machine learning methods, unsupervised learning methods require less feature engineering efforts in obtaining features consequently saving the time and the labor. In \cite{YLei:IEEETIE2016}, a two-stage unsupervised learning method was proposed to conduct diagnostic analysis of machine faults. In addition to fault diagnosis, \emph{anomaly detection} (or outlier detection) is to identify data objects that do not comply with an expected pattern as given. In \cite{ZZheng:TII18}, a deep learning based method was proposed to detect electric theft via anomaly detection of electricity consumption data in smart grids.

\emph{(3) Predictive analytics}

Predictive analytics mainly utilizes historical data to anticipate the trends of data (i.e., what will occur in the future). In\cite{wu2017comparative}, a random forests (RFs) based method was proposed to predict the tool (machine) wear in manufacturing cycle. It is also shown in \cite{wu2017comparative} that RFs method outperforms ANN and SVMs in terms of prediction accuracy. One of challenges in data analytics of MIoT data is the imbalanced number of negative and postive samples \cite{lade2017manufacturing}. The study of \cite{kim2017imbalanced} proposed a cost-sensitive decision tree ensemble algorithm to address this issue. Extensive experimental results show that the proposed method outperforms other existing baseline methods. Moreover, in \cite{RRen:IEEETCybernetics2017}, a deep-learning based method was proposed to predict product surface defects. In addition, consumer behaviour prediction plays an important role in manufacturing business stage, e.g., to improve the consumers' purchase decision-makeing predictions. In \cite{YZuo:2016}, a Bayesian network based approach was proposed to predict the customer purchase behaviour. In particular, the analysis is based on massive RFID data, which was collected through RFID tags attached at customers.

\emph{(4) Prescriptive analytics}

Prescriptive analytics extends the results of descriptive, diagnostic and predictive analytics to make right decisions in order to achieve predicted outcomes (i.e., what should we do to achieve the goal?). The prescriptive methods typically include \emph{simulation}, \emph{decision-making}, \emph{optimization} and \emph{reinforcement learning algorithms}. In particular, in \cite{Gerlach:processes2015}, a conceptual design approach was proposed to simulate the configuration and procedural training in a bio-ethanol plant. The study of \cite{Mourtzis:CIRP2016} presents a novel method for manufacturing-networks design via intelligent decision-making on selecting suppliers to fulfill the requirements of frugal innovation. In \cite{kluczek2016application}, an analytic hierarchy process (AHP) based method was proposed to evaluate manufacturing sustainability performance. Moreover, in \cite{Drakaki:App2017}, a novel method with the integration of Timed Colored Petri Nets (CTPNs) and reinforcement learning (RL) was proposed to solve the problem of manufacturing scheduling.

\begin{table*}[t]
\centering
\caption{Classification of data analytics approaches in MIoT}
\vspace{0.2cm}
\label{tab:class}
\small
\begin{tabular}{m{1.5cm}|m{2.6cm}|m{5.9cm}|m{4.2cm}|m{2.4cm}}
\hline
             & \begin{center}\bf Questions\end{center}            & \begin{center}\bf Approaches                                                                                                                                     \end{center}   & \begin{center}\bf Applications                                                                                                                 \end{center}   & \begin{center}\bf References                                                                                                                 \end{center} \\\hline\hline

Descriptive  & What happened?       & \begin{tabular}[c]{@{}l@{}} \tabitem Association rule mining\\ \tabitem Clustering, sequential pattern mining\\ \tabitem Querying, statistic reporting \\ \tabitem Data visualization\end{tabular}              & \begin{tabular}[c]{@{}l@{}} Misbehaviour pattern capturing \\Product status checking\\ Benchmark analysis \end{tabular} & \cite{zhang2015:real} \cite{Zhong:IJAMT2016} \cite{zuo2018internet} \cite{MOLKADANIELSEN:2018}\\\hline

Diagnostic   & Why it happened?     & \begin{tabular}[c]{@{}l@{}}\tabitem Reasoning\\ \tabitem Bayesian analysis\\ \end{tabular}                                                          & \begin{tabular}[c]{@{}l@{}}  Fault diagnosis \\ Root-cause analysis of failure  \\ Anomaly detection \end{tabular} & \cite{AZADEH:2013} \cite{HWang:OCEANS2016} \cite{Wuest:2014} \cite{YLei:IEEETIE2016} \cite{ZZheng:TII18} \\\hline

Predictive   & What might happen in the future?   & \begin{tabular}[c]{@{}l@{}}\tabitem Classification, regression\\ \tabitem Machine learning (supervised\\\hspace{0.2cm} /unsupervised)\\ \tabitem Deep learning\end{tabular} & \begin{tabular}[c]{@{}l@{}}Trajectory prediction\\ Consumer behaviour prediction\\ Device maintenance prediction\end{tabular} & \cite{wu2017comparative} \cite{kim2017imbalanced} \cite{RRen:IEEETCybernetics2017} \cite{YZuo:2016}\\\hline

Prescriptive & What should be done? & \begin{tabular}[c]{@{}l@{}}\tabitem Simulation\\ \tabitem Optimization\\ \tabitem Reinforcement learning (e.g., Q-Learning)\\ \tabitem Decision making: e.g., Analytic Hierarchy 
\\\hspace{0.22cm} Process (AHP), The Technique for Order 
\\\hspace{0.22cm} of Preference by Similarity to Ideal
\\\hspace{0.22cm} Solution (TOPSIS) \end{tabular}                                  & \begin{tabular}[c]{@{}l@{}}System resilience\\ System reliability\\ System optimization\end{tabular}   & \cite{Gerlach:processes2015} \cite{Mourtzis:CIRP2016} \cite{kluczek2016application} \cite{Drakaki:App2017}
\\\hline
\end{tabular}
\end{table*}

Table \ref{tab:class} summarizes data analytics methods used for MIoT. We categorize them into four types according to different levels in terms of complexity and extracted values. Moreover, we also enumerate representative data analytics methods in each category. In addition, we also list representative application cases in each category.

\subsubsection{Data visualization in MIoT}

In addition to the aforementioned data analytics, data visualization is also an important tool in MIoT data. Effective data visualization procedure can help to extract and interpret the informative values from complex and high-dimensional MIoT data \cite{telea:2014}. Typical data visualization methods include information visualization, exploratory data analysis, statistic plots. The typical quantitative messages that are conveyed by data visualization include: time-series, ranking, frequency distribution, deviation, correlation, part-to-whole, geographic \cite{Frits:2003}. The basic data visualization techniques include: 1) various statistic plots (e.g., bar chart, histogram, pie diagram, scatter plots), 2) word clouds of text data, 3) correlation coefficient matrices/functions, 4) network/graph diagrams of non-structural data, 5) heat map of geographic data.

Typical data visualization toolboxes include Matlab plot (\url{https://it.mathworks.com/help/matlab/ref/plot.html}), gnuplot (\url{http://www.gnuplot.info/}), Python's Seaborn (\url{https://seaborn.pydata.org/}), Pandas plot (\url{https://pandas.pydata.org/}), Matplotlib (\url{https://matplotlib.org/}). Moreover, web-based visualization tools have also been wide used. Representative web-based data visualization tools include Tableau (\url{https://www.tableau.com/}), Plotly (\url{https://plot.ly/}), Sisense (\url{https://www.sisense.com/}), D3.js (\url{https://d3js.org/}).


\subsection{Case studies}
\label{subsec:case}



\begin{figure*}[t]
\centering
\subfigure[System Prototype]{
\includegraphics[width=7.0cm]{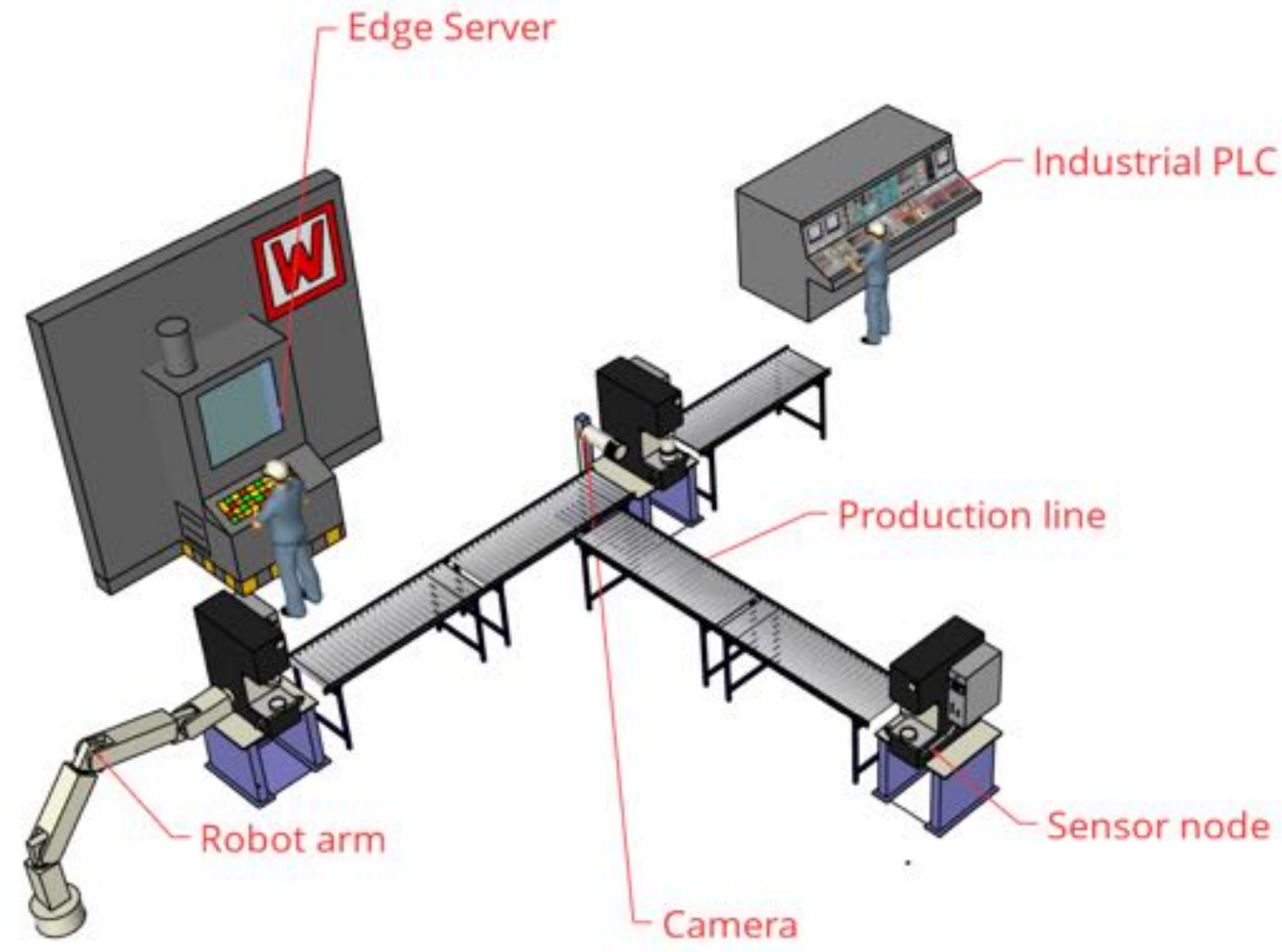}
\label{fig:sys-prototype}}
\subfigure[Realistic deployment of system prototype]{
\includegraphics[width=5.8cm]{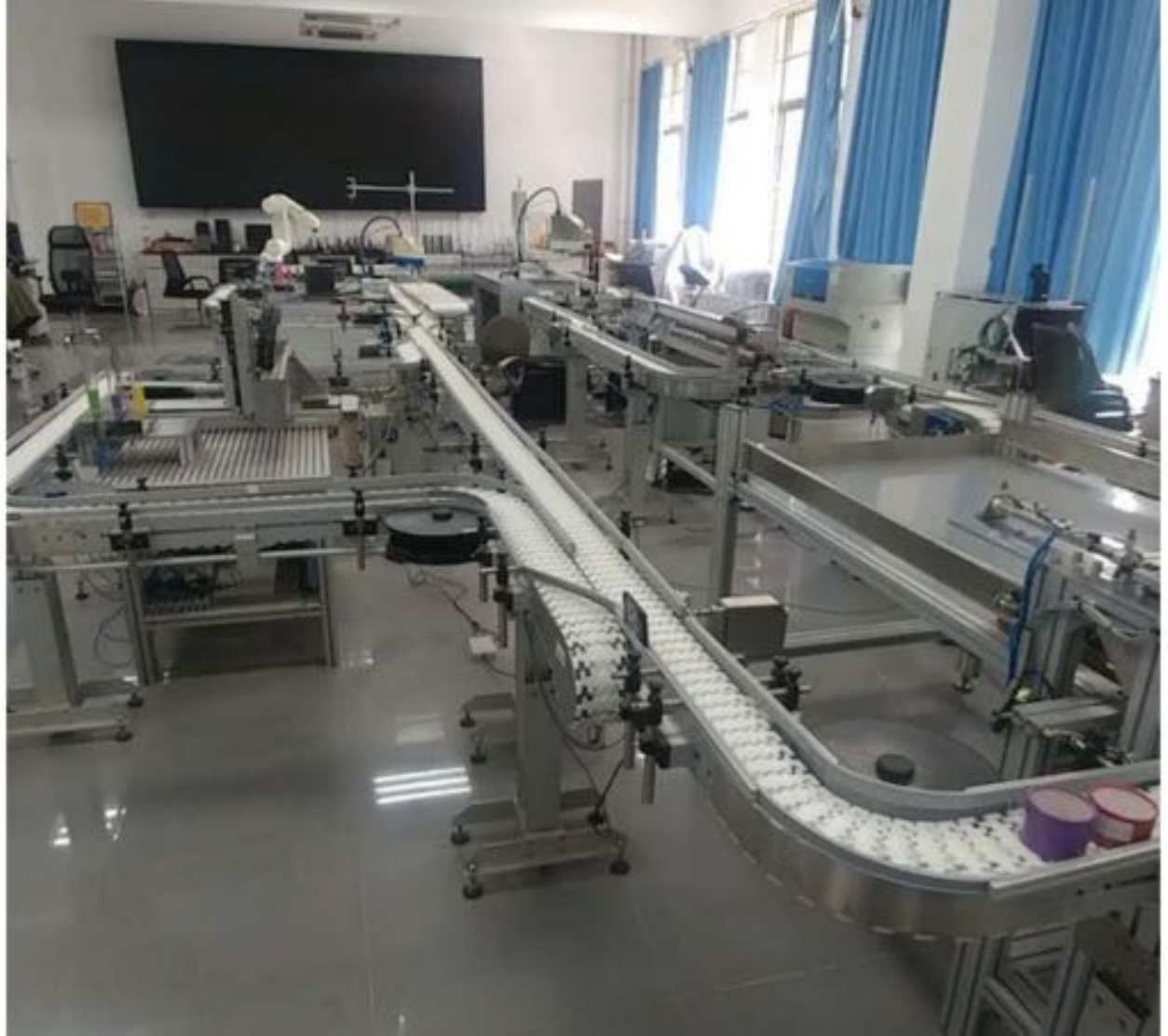}
\label{fig:deployment}}
\caption{Case study for distributed computing models for MIoT}
\label{fig:case}
\end{figure*}

To demonstrate the feasibility of distributed computing models in MIoT, we developed a system prototype. Figure \ref{fig:sys-prototype} shows that the system framework consists of a production line, industrial devices and computing units. In particular, the production line consists of various manufacturing devices, instruments, sensors, actuators and robot arms, all of which are connected through wired or wireless links consequently forming the MIoT. In addition to the production line and industrial devices, there are a number computing units supporting diverse data processing tasks. For example, edge computing servers with equipped with embedded computers are deployed in the proximity to MIoT.  Moreover, the computing-intensive tasks may be uploaded to the remote cloud servers while the latency-sensitive tasks may be processed at edge servers. 

In the computing perspective, we develop a distributed computing platform with the orchestration of remote cloud computing and local edge computing. In particular, we deploy Xen hypervisor at remote cloud servers and Docker container at edge servers. On top of virtual machines, we further utilize Hadoop distributed computing platforms to support big data processing tasks. In order to coordinate the edge and cloud computing tasks, we design and implement a hybrid edge/cloud computing framework (details can be referred to the work \cite{XLi:TII19}).

Figure \ref{fig:deployment} gives the realistic prototype of a printed circuit board (PCB) production line based on our proposed system framework. This production line consists of conveyor belts, product feeding machines, robot arms, sensors and cameras. We choose industrial WLANs as the wired connections and 6LoWPAN as the wireless connections. In addition, we adopt 4 edge servers, each of which has the identical configurations: a single-board computer with a quad-core Broadcom BCM2837 CPU, 1GB memory and 64GB SSD storage. Furthermore, there is a remote cloud server (i.e., IBM X3650 M3) with 2 Intel Xeon Processors, 24 GB memory and 1TB SSD storage.  

We then evaluate the performance of the proposed hybrid edge/cloud computing framework on top of the prototype. In particular, we consider a pure cloud computing framework and a pure edge computing framework as baseline models. Moreover, image recognition tasks with varied image size were chosen to be executed at edge and cloud servers. We further adopt OpenCV frameworks on both edge and cloud servers to support the image recognition tasks.

Table \ref{tab:performance} shows the latency values of three computing frameworks versus varied image sizes. In particular, the latency is calculated via averaging results with 100 images, each with the same image size (e.g., 10 MB). It is shown in Table \ref{tab:performance} that the average latency is increased with the increased image size; this effect may owe to the increased computational complexity of image recognition algorithms with the increased image size. We also observe from Table \ref{tab:performance} that the proposed hybrid cloud and edge scheme outperforms pure cloud computing scheme and pure edge computing scheme with larger image size (e.g., 16 MB, 18 MB and 20 MB). It can be explained as follows: 1) pure cloud computing has the strength in processing large images while suffering from the long end-to-end latency; 2) pure edge computing scheme can complete the computing tasks with smaller image size (e.g., 12 MB) and achieve the short end-to-end latency due to the deployment proximity; 3) hybrid edge/cloud computing scheme can not only exploit the strength of cloud computing to process the complicated tasks but also harness the benefit of edge computing in short latency, consequently obtaining the better performance in the cases with larger image size.



\begin{table*}[t]
\centering
\caption{Performance evaluation}
\vspace{0.2cm}
\renewcommand{\arraystretch}{2.0}
\footnotesize
\begin{tabular}{|c|c|c|c|c|c|c|}
\hline
    & 10 MB          & 12 MB          & 14 MB          & 16 MB          & 18 MB          & 20 MB          \\ \hline
Cloud Computing Only (second)  & 1.20          & 1.48          & 1.67          & 1.82          & 2.08          & 2.45          \\ \hline
Edge Computing Only (second)   & \textbf{0.61} & \textbf{0.86} & \textbf{0.97} & 1.15          & 1.26          & 1.43          \\ \hline
Hybrid Cloud and Edge (second) & 0.75          & 0.93          & 0.98          & \textbf{0.86} & \textbf{0.97} & \textbf{0.96} \\ \hline
\end{tabular}
\label{tab:performance}
\end{table*}

\section{Future research directions}
\label{sec:open}

\begin{figure*}[t]
\centering
\includegraphics[width=15cm]{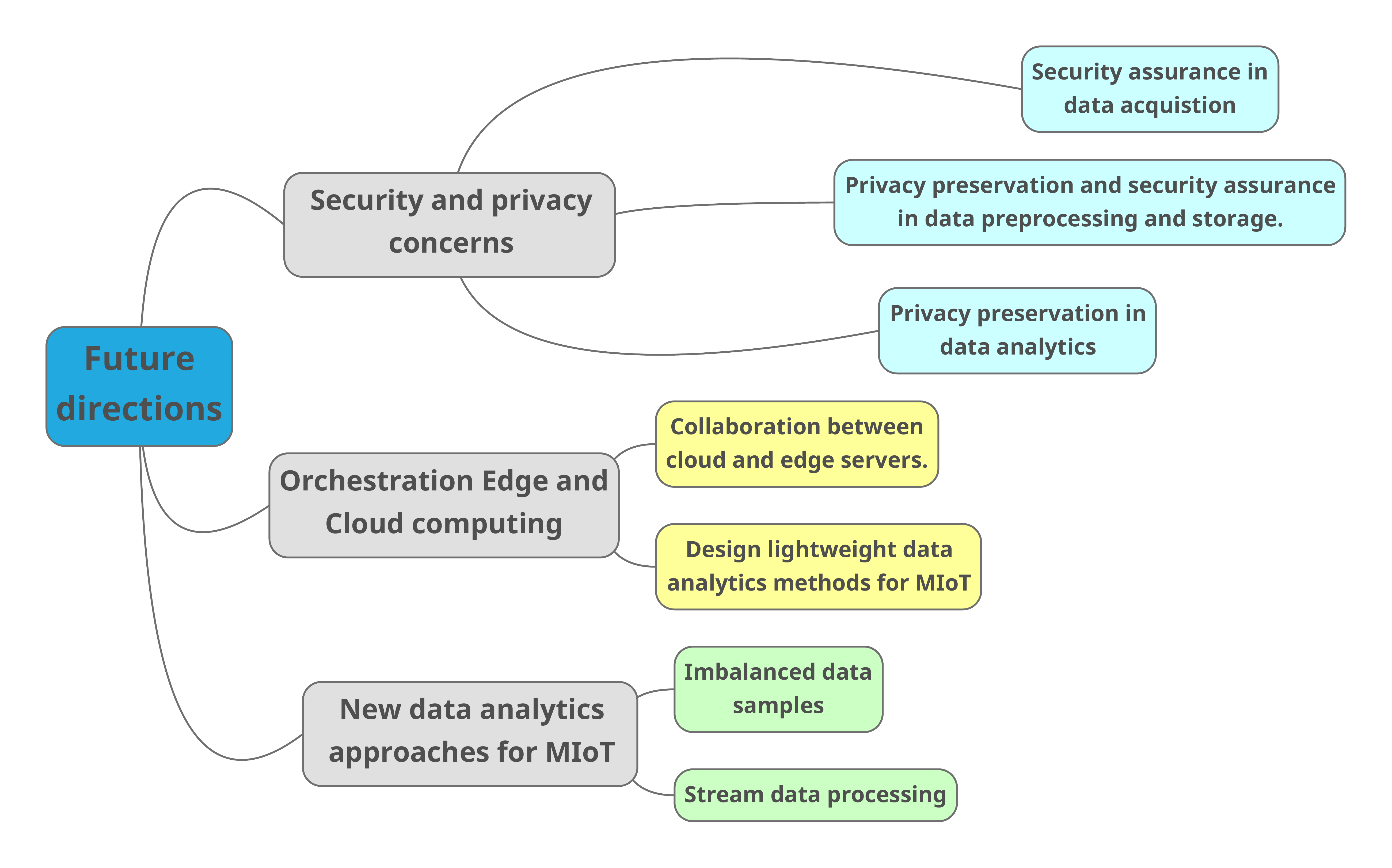}
\caption{Future directions in big data analytics of MIoT}
\label{fig:future}
\end{figure*}

In this section, we discuss open issues as well as future directions in big data analytics for MIoT. Figure \ref{fig:future} summarizes the future directions in big data analytics in MIoT.

\subsection{Security and Privacy Concerns}
\label{subsec:sec_privacy}

Privacy and security are becoming an arising challenge of big data analytics for MIoT. Privacy concerns the proper utilization of the data with the preservation of enterprise private information, whereas  security is to ensure data confidentiality, integrity and availability \cite{XWang:TBD2018}. We next summarize the research issues related to privacy and security in big data analytics for MIoT.

\begin{itemize}

\item \emph{Security assurance in data acquisition}. The proliferation of wireless connections in manufacturing industry results in the challenges in security assurance during data acquisition because of the openness of wireless medium susceptible to malicious attacks like passive eavesdropping attacks \cite{lxr:Sensors2018}. The typical countermeasure is to apply encryption schemes in wireless networks \cite{Hennebert:IoTJ2014}. However, it may not be feasible to apply cryptography-based techniques in all IoT networks due to the following constraints: the inferior computational capability and the limited battery power of some smart objects like RFID and sensors. Therefore, new protection schemes without strong computational complexity and high energy consumption shall be developed for MIoT in the future. Blockchain, featured with security and reliability, can potentially improve the security and reliability of MIoT \cite{hndai:blockchain-iot2019}.

\item \emph{Privacy preservation and security assurance in data preprocessing and storage.} After data acquisition, MIoT data will be preprocessed and stored locally (at servers of factories or other departments) or remotely (at remote cloud servers) \cite{wang2015cloud}. However, the distribution of MIoT data throughout the enterprise consisting of multiple manufacturing sites across different regions often results in the vulnerability to various malicious attacks from insiders and outsiders of the enterprise. It is challenging to offer a solution against malicious attacks. There are several possible directions in solving this issue: 1) Proper key management \cite{Esposito:IEEECloudComp2016} including proper key distribution and key validation period, 2) authentication mechanism including accessing control of files and data records, 3) traceability of data accessing allowing any data accessing or modification to be identifiable so that the malicious behaviours can be avoided or revoked. 

\item \emph{Privacy preservation in data analytics.} In order to protect data privacy, the data is often encrypted and stored at a server (or at a cloud). Before data analytics, the data needs to be decrypted. However, the decryption process is often time-consuming consequently resulting in the inefficiency of data analytics in MIoT. How to design a privacy-preservation scheme of balancing the efficiency and privacy becomes a challenge \cite{xkxiao:icde2018,BABAR:2019}. 

\end{itemize}

\subsection{Edge Computing for big data analytics in MIoT}

The integration of cloud computing with manufacturing brings the opportunities in saving the capital investments of information and communication technologies (ICT), providing flexibility of ICT resources to small and medium enterprises \cite{wang2015cloud,Esposito:IEEECloudComp2016}. However, there are also limitations with cloud computing such as high latency, performance bottleneck, single-point-to-failure and privacy leakage \cite{HLiu:IEEESys17}. Recently, mobile edge computing (or fog computing) has become a new complement to cloud computing by offloading both \emph{computational and storage tasks} from remote cloud servers to local edge servers \cite{Tran:ComMag17,WU:JMS2017,XWang:ComMag17}. In this manner, the computing-intensive and delay-tolerant tasks will be executed at remote cloud servers while the delay-critical and computing less-intensive tasks will be offloaded to edge servers. As a result, the real-time tasks like sensing, monitoring and controlling can be enabled in the proximity to factories and enterprises. The case study in Section \ref{subsec:case} also demonstrates the effectiveness of hybrid edge and cloud computing in MIoT.

However, there are many challenges in edge computing for big data analytics in MIoT. 
\begin{itemize}
\item \emph{Collaboration between cloud and edge servers.} There are diversity of computing resources in manufacturing networks. For example, remote cloud servers usually have superior computing capability than local edge servers while there is a longer delay to upload the tasks to the remote cloud servers than to upload the tasks to the local edge servers Therefore, it is necessary to determine how to allocate the computational tasks at cloud servers or at edge servers. For example, the computing intensive and delay-tolerant tasks should be uploaded to remote cloud servers while the computing less-intensive and delay-critical tasks can be executed locally at edge servers. In this sense, edge servers can be deployed within factories and remote clouds can be deployed outside factories (even if they can be provided by third parties). To the best of our knowledge, there are few studies on investigating collaboration between cloud and edge servers, especially in the whole manufacturing network. In the future, research efforts should be done in allocating and coordinating various computing resources distributed in cloud and edge servers in manufacturing.

\item \emph{Design lightweight data analytics methods for MIoT.} Many data analytics tasks that are delay-critical should be executed locally at edge servers (or at manufacturing devices). However, due to the resource limitation of edge severs, the conventional data analytics methods might be too complicated to be executed at edge servers. Therefore, the models of the data analytics methods need to be trained at remote cloud servers first and be transferred at local edge servers. However, it can result in huge communication cost to transmit this model from the remote cloud servers to the edge servers. For example, the study of \cite{YLin:ICLR2018} shows that AlexNet (i.e., a typical deep learning method) has the model size of 240MB, which is so large that it can cause extra delay from the cloud server to the edge server. Therefore, it is necessary to design lightweight data analytics schemes which can be deployed locally at edge servers approximate to users \cite{CLeng:AAAI18}.
\end{itemize}  

\subsection{New data analytics methods for MIoT data}
\label{subsec:data_analysis}

Although a lot of efforts have been done in developing data analytics methods for MIoT data, there are still many open research issues in this area. 

\begin{itemize}

\item \emph{Imbalanced data samples}. Different from data analytics in traditional fields (e.g., commercial database systems), manufacturing data has the imbalanced number of data samples between positive and negative samples. For example, it is shown in \cite{lade2017manufacturing} that the ratio of positive samples to negative samples (vice versa) can be 99,000,000 to 1. It is challenging to apply conventional data analytics methods to analyse the imbalanced dataset. Therefore, new data analytics methods should be developed to solve this issue. To the best of our knowledge, there are few studies \cite{kim2017imbalanced} proposed to address this issue.

\item \emph{Stream data processing}. In MIoT, there is a tremendous volume of real-time data generated (e.g., sensory data from industrial wireless sensor networks) \cite{XWang:TCCS18}. It is impossible to store and process the entire data in the memory of computers. Consequently, the conventional methods requiring saving the whole data sets in memory cannot work in this scenario. It is challenging to analyse the massive data-stream of MIoT. It is worthwhile to investigate new data analytics approaches to process the data-stream of MIoT.

\end{itemize}

\section{Conclusion}
\label{sec:conc}


This paper presents an in-depth survey on big data analytics in manufacturing Internet of Things (MIoT). This paper first presents a life cycle of big data analytics in MIoT and discusses the necessities as well as challenges of big data analytics in MIoT. Then, the enabling technologies of big data analytics in MIoT are summarized according to three phases in the life cycle of big data analytics: data acquisition, data preprocessing and storage, and data analytics. Moreover, this paper also outlines the future directions and discusses the open research issues. We believe big data analytics will play an important role in promoting manufacturing industry to evolve into smart manufacturing in the foreseeable future. 

\section*{Disclosure statement}
The authors declare that they have no potential conflict of interest.

\section*{Funding}
The research of Hong-Ning Dai and Hao Wang is supported by Macao Science and Technology Development Fund under Grant No. 0026/2018/A1, National Natural Science Foundation of China (NFSC) under Grant No. 61672170, NSFC-Guangdong Joint Fund under Grant No. U1401251, and Science and Technology Program of Guangzhou under Grant No. 201807010058. Guangquan Xu's work is supported by the State Key Development Program of China (No. 2017YFE0111900), National Science Foundation of China (No. 61572355, U1736115). Jiafu Wan's work is supported by Science and Technology Program of Guangzhou (No. 201802030005), Guangdong Province Key Areas R \& D Program (No. 2019B090919002). Muhammad Imran's work is supported by the Deanship of Scientific Research, King Saud University through research group number RG-1435-051.

\bibliographystyle{tfcad}
\bibliography{reference}

\end{document}